# The Superluminal Tunneling Story

*Horst Aichmann* [1] *and Günter Nimtz* [2]

[1] Zur Bitz 1, 61231 Bad Nauheim/Germany
   Worked in microwave technology at HP / Agilent Technologies for 25 years
[2] II. Physikalisches Institut, Universität zu Köln,
   Zülpicher Str. 77, 50937 Köln/ Germany

**Abstract**

Since 1992 experimental evidence of superluminal (faster than light, FTL) signals are causing much excitement in the physics community and in the media. Superluminal signal velocity and zero time tunneling was first observed in an analog tunneling experiment with microwaves [1]. Recently, the conjectured zero time of electron tunneling [2-5] was claimed to be observed in ionizing helium [6]. The FTL signal velocity was reproduced with infrared light and with various barriers in several laboratories worldwide [7]. Remarkable, it was shown that the tunneling time is a universal quantity for elastic and for electromagnetic fields [8,9,10]. Many theoretical physicists predicted this FTL nature of the tunneling process. However, even with this background many members of the physics community did not accept the superluminal signal velocity interpretation of the experimental results and they also ignored the universal tunneling time. The predicted and measured zero tunneling time inside a barrier was taken as a fantastic nonsense. They tried with passion to prove the interpretation to be incorrect, even though the experiments have been quantitatively described by several theoretical approaches. The antagonists were scared that a superluminal signal velocity would violate Einstein causality and time machines would become possible. In fact superluminal signals were shown to confront the Einstein causality, but they don't violate primitive causality. Tunneling is not a classical process, and as this not described by the special theory of relativity (STR).

First we want to resume some historical investigations about the tunneling process and after that we introduce and discuss some of the not-FTL interpretations of the measurements. One of the most boisterous protagonists on this subject, Herbert Winful, claimed in about a dozen of papers to have ***resolved the mystery of apparent superluminality by an evanescent cavity*** and called ***the Hartman effect as the heart of the tunneling time conundrum*** [11]. Similar studies are published in textbooks and in leading physical journals, in spite of the fact that the condemned interpretation and calculations quantitatively describe the experimental data. Zero time tunneling and virtual particles are accepted in the microcosm, however, not in the macrocosm by the STR community.
A brief explanation of the misleading theoretical approaches against FTL signals due to the tunneling process is presented, eventually.

**Background**
Einstein elaborated in the Special Theory of Relativity (STR) in 1905 that the utmost velocity of waves and particles is given by the light velocity c in vacuum. Investigations on this velocity followed by Sommerfeld and Brillouin. For this reason they studied the interaction of an electromagnetic wave with the Lorenz-Lorentz dipole oscillator [12]. This classical interaction is dissipative and lossy. The calculations were carried out with signals of unlimited frequency bands. They have discussed and defined the relevant velocity in the STR. Brillouin was aware that a physical signal begins and ends gradually, as mentioned in his textbook [12]. Sommerfeld's and Brillouin's classical model is not appropriate to describe the tunneling process and the evanescent modes, where wave packets have an imaginary wave number and an elastic barrier interaction. In addition, physical signals are frequency band limited. An unlimited frequency band demands an infinite energy, since a field does only exist, if it has at least one quantum at the frequency in question.

On the other hand Friedrich Hund conjectured non classical barrier penetration in the case of the inversion motion in ammonia in 1927 [13]. At the same time tunneling was studied by Nordheim applying the Schrödinger equation. Oppenheimer performed a correct calculation of the rate of ionization of hydrogen by an external field in 1928. Gamow and independently Gurney and Condon applied the tunneling phenomenon to explain the range of alpha decay rates of radioactive nuclei. Later in the sixties tunneling was investigated with respect to solid state physics, where electrons tunnel thin insulating layers sandwiched in superconductors and electrons traverse the band gap in semiconductor Esaki diodes.

Several theoretical studies concluded that tunneling takes place inside a barrier in zero time [2-5,14]. Other studies and text-books denied fiercely such results up to now, see for example [5,11,16,17]. Remarkable, as well these authors did not note the observed universal interaction time $\tau_i$. This time is given by the relation $\tau_i \approx h/E = 1/\nu$, where h is the Planck constant, E the wave packet energy, and $\nu$ the wave packet center frequency [8,9,10].

We present and discuss four examples out of many publications, which have apparently shown that the measured superluminal signals have not been superluminal. The superluminal velocity results due to zero tunneling time and the short front barrier interaction time as was calculated already by Hartman in 1962. Remarkable, within the universal barrier front interaction time the reflected and the transmitted wave packet learned about the barrier's height and length. This time is shorter than the corresponding vacuum traversal time would be.

**Examples**

I. In the STR textbook by Sexel and Urbantke [16] one reads:

*…….In these cases (of claimed FTL), dispersion is so pronounced that the concept of wave packet becomes rather meaningless, as an initial packet gets completely deformed and unsuitable for perfect signal transmission………*

We think, the authors did not read the papers with the published superluminal signals displayed in Figs. 1; 2, for instance.

Another unphysical statement is found on pages 24-26 in this textbook:

*Since perfect signals are to be regarded always as a kind of discontinuity…….*

Such a perfect signal demands an infinite frequency band and thus an infinite energy.

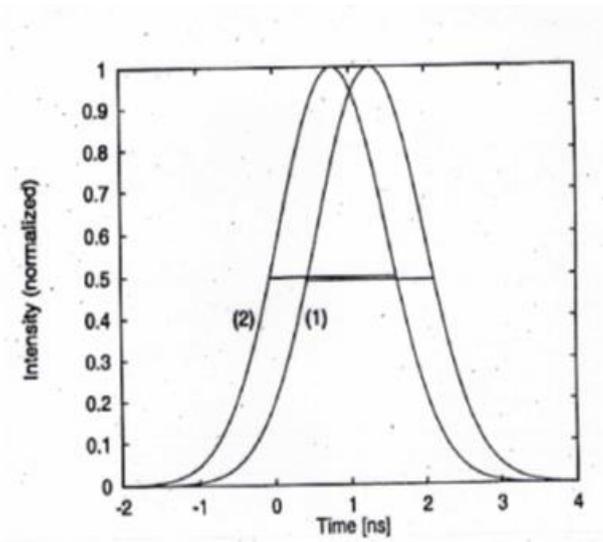

Fig.1 Two microwave pulses (8.2 GHz), (1) travelling through air and (2) through a barrier, the latter traversed the same distance at 4.7 c. There is no reshaping and the signal i.e. the half width is the same. In todays optical communication systems the time halfwidth of the pulse represents the number of consecutive ones ( or zeros ) in a data stream see also Figs. 2, 5.

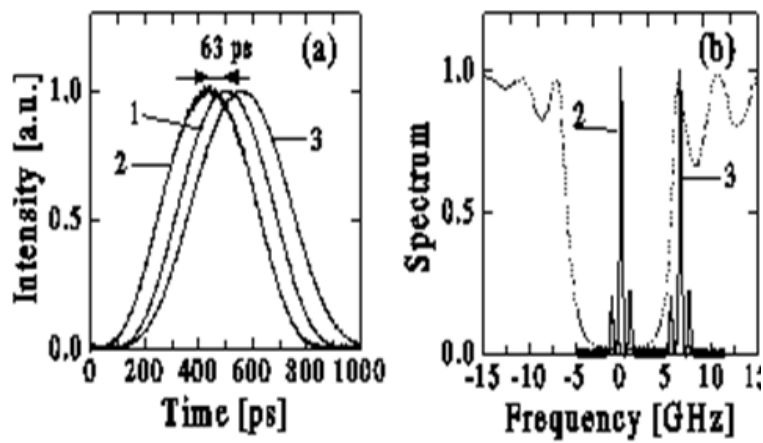

Fig. 2. An example of a tunneled digital infrared signal. 1 represents the tunneled, 2 the vacuum travelled signal at 5 c. 3 represent the signal travelling slowly outside the tunneling frequency regime. The digital signal is given by its halfwidth [18].

The presented Eq. (2.34) for the front velocity [16], is irrelevant in the case of purely imaginary wave numbers. The front velocity is defined by $v_F = \lim_{\omega\to\infty} \omega/k$, this quantity is neither existent in physical signals nor in single photons, electrons or phonons, since only the energy quantum $\hbar\omega$ is measurable since Planck's discovery in 1900.

II. Moses Fayngold wrote in the textbook Special Relativity and Motions Faster than Light:

**_Thus, as we take a close look at the fast light, there is no room left for the notion of a superluminal communication or superluminal energy transfer (Page 223)._**

According to this conclusion by Fayngold, the digital microwave signals as well as the infrared signals shown in Figs.1; 2 would have been measured without energy.

Another statement by Fayngold:

**_Because all observable properties of material objects are real, the appearance of the imaginary values in the theory indicates that corresponding quantities cannot be measured. But what cannot, in principle, be observed does not exist. In other words there cannot be any superluminal particles._**

It is correct that a particle (or wave packet) cannot be measured tunneling inside a barrier, however, outside the barrier as in the case of radioactivity or in the case of the signals as demonstrated in Figs.1;2 and explained in Fig. 6.

The STR is also confronted by the fact that the energy relation $E^2 = (\hbar k c)^2 + m^2 c^4$ is not fulfilled, since the wave number k of evanescent and tunneling modes is imaginary. This property and the consequences of evanescent modes are discussed e.g. in [10,19,20].

III. Raymond Chiao et al.: Tunneling Single Photons and the Tortoise Race:

Raymond Chiao et al. have explained their superluminal single photon measurements in Fig. 4 by a superluminal velocity of the center of mass. This was illustrated by a race of tortoises. One ensemble of tortoises was travelling along vacuum and another one the same distance through a barrier. As the tunneled one was much smaller due to reflection only the center of mass of a strongly reshaped signal travelled superluminal, whereas the front of both arrived with c. The authors did not realize that the tunneled signal was smaller but should have still the same shape – from forte to piano, with the same halfwidth as demonstrated in the sketch of Fig. 5.

At the same time they have published the coincidence profiles of the single photon experiment in Fig. 3 of Ref. [21] . In Fig. 3 we display a sketch of the identically measured coincidences.

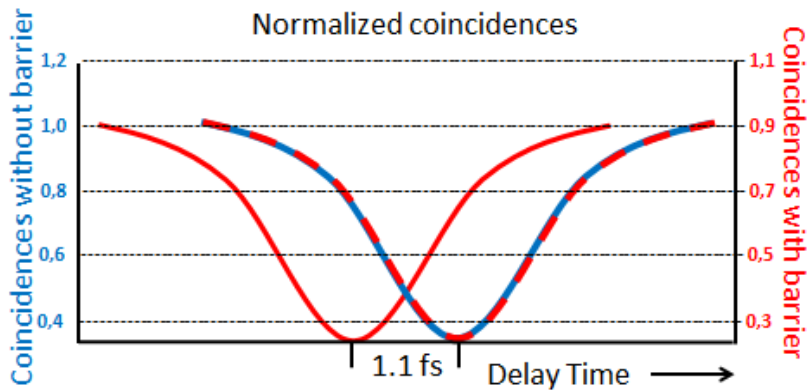

Fig. 3. A sketch of the measured coincidences in [21]. The fast tunneled photon coincidence (red) displays an identical behavior as the blue air traveled photons, seen by the shifted broken red line.

In the tortoise race [22], the front velocity was said to be the decisive velocity. However, comparing the measured data of Fig.3 published in PRL with the humorous tortoise sketch in Scientific American published at the same time, one can see that the measured front of the tunneled photons have won the race. Namely, the measured single photon coincidence profile is exactly the same for the tunneled as well as for the vacuum traveled photons. Thus the coincidence front of the tunneled single photons is 1.1 fs faster than those traveled through vacuum.

A front velocity has no physical significance as explained above. However, Chiao and Steinberg have an opposite opinion and write in the last sentence of their review [21]

***….we believe that their point of view (no central significance of the front velocity for signals: Nimtz and Heitmann [23] ) is fundamentally incorrect.***

Their single photon experiments were performed with frequency bandwidths of the photons $< 10^{-1}$ $\nu_0$, where $\nu_0$ is the photon frequency. Thus the photons have neither an infinite frequency bandwidth nor a non-analytical behavior.

Even a single photon or electron is signaling that they have been emitted, this information becomes particularly obvious in the case of radio activity.

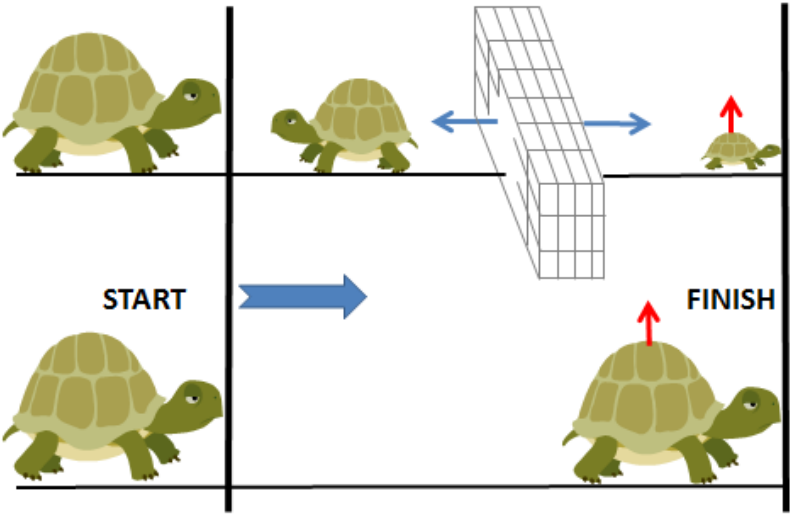

Fig. 4. The tortoise race, the tunneled one (photon) is deformed and only its center of mass was faster than that of the not tunneled photons.

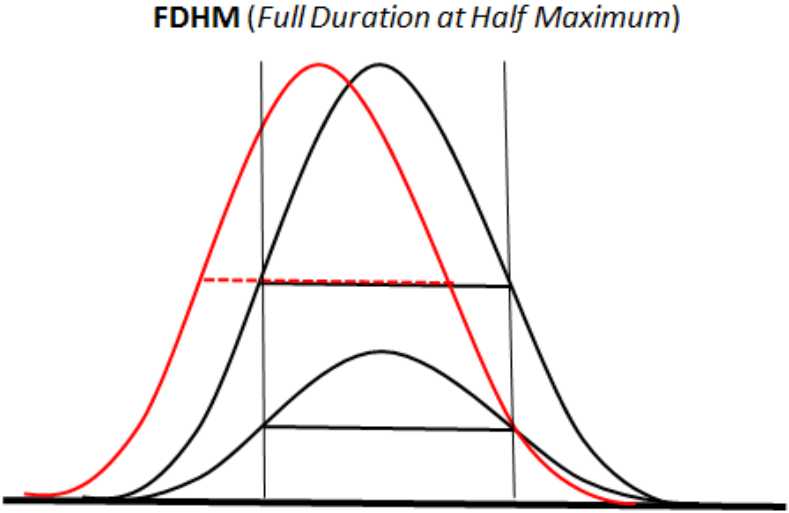

Fig. 5. Intensity vs time of a fast and a slow digit. The attenuated signal has still the same information, i.e. the same FDHM.

    **IV.** Steinberg's special tunneling train:

Professor Steinberg of the University of Toronto has also stated that Nimtz has not demonstrated Einstein causality violation (which would be implied by transmitting information faster than light). In a New Scientist article, he uses the analogy of a train traveling from Chicago to New York, but dropping off train wagons at each station along the way, so that the center of the train moves forward at each stop; in this way, the speed of the center of the train exceeds the speed of any of the individual wagons.

However, tunneling is an elastic process and after reflection took place at the barrier front any transmitted wave packet of electrons or photons is not attenuated, thus no wagon is dropped off between Chicago and New York.

**V.** Winful's tunneling barrier cavity:

Herbert Winful has vehemently explained the tunneling transmission and reflection time by the decay time of a cavity, which represents a tunneling barrier. According to his studies after a meeting on Quantum Optics, Kavli Institute for Theoretical Physics, Santa Barbara, July 2002 he published his tunneling time explanation frequently in leading physical journals. The first paper appeared in Optics Express (V. 10, p. 1491 (2002) with the following abstract [11].

***We show that the anomalously delay times observed in barrier tunneling have their origin in energy storage and its subsequent release. The observed group delay is proportional to the energy stored. This delay is not a propagation delay and should not be linked to a velocity since evanescent waves do not propagate. The Hartman effect, in which the group delay becomes independent of thickness for opaque barriers, is shown to be a consequence of the saturation of stored energy with barrier length.***

Winful argues that the train Steinberg analogy is a variant of the "reshaping argument" for superluminal tunneling velocities, but he continues to say that this argument is not actually supported by experiment or simulations, which show that the transmitted pulse has the same length and shape as the incident pulse. Instead, Winful argues that the group delay in tunneling is not actually the transit time for the pulse (whose spatial length must be greater than the barrier length in order for its spectrum to be narrow enough to allow tunneling, (this is correct due to avoid dispersion effects, which was show by Hartman 1962)), but is instead the lifetime of the energy stored in a standing wave which builds up inside the barrier. Since the stored energy in the barrier is less than the energy stored in a barrier-free region of the same length due to destructive interference, the group delay for the energy to escape the barrier region is shorter than it would be in free space, which according to Winful is the explanation for apparently superluminal tunneling.

There are theoretical and experimental facts, which are not in agreement with the cavity model:

1. the measured tunneling time is by about one order of magnitude shorter than the time to build a standing wave in a cavity.

2. the established universal time is given by the reciprocal of the wave packets energy (see above). This property is not given by the cavity approach, a Q value is not related to the cavity frequency.
3. the calculated traversal times, which Winful calls cavity decay times are about a factor 0.1 shorter than the measured and calculated transversal times by various phase time approaches.
4. Evanescent and tunneling modes are not measurable. They are virtual particles as sketched in Fig. 6.

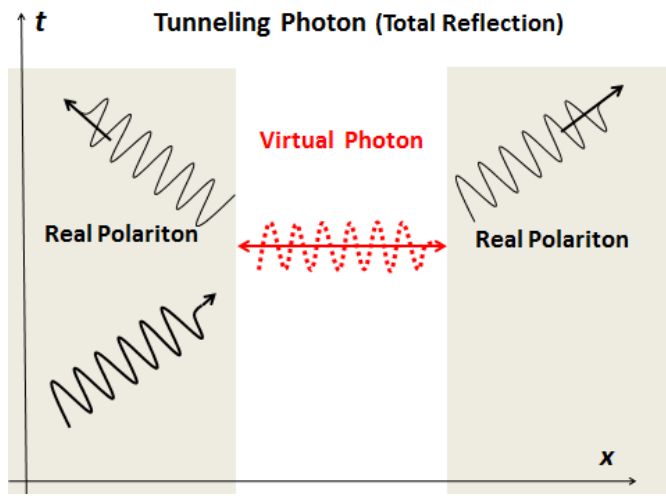

Fig. 6. A virtual photon inside the barrier in the case of total frustrated internal reflection (TFIR).

**Summing-up:**
Many physicists see the FTL interpretation of tunneling like the poet Morgenstern [24] : What cannot be must not be.
In order to prove the FTL interpretation to be incorrect, unphysical assumptions are proposed. Usually the STR crusaders assume signals to have infinite frequency bands, which results in signal reshaping and in a luminal front velocity due to the dispersion of a barrier. Another incorrect assumption is a signal to have a point-like time. A signal and thus the information are always given by the product of time duration and frequency bandwidth. For example even the single photons had a time duration of the order of ≈100 fs and a bandwidth of the order of 10 THz at a center frequency of 427 THz in Ref. [21]. There is no front velocity as was claimed by Chiao et al. [15,21,22].
On the other hand, it was shown that due to the finite time duration a superluminal signal may begin in the past but it will always end in the future [25]. This is in consequence of the dispersion of any tunneling barrier and excludes the design of a time machine with FTL signal carriers [25].

Incidentally, a wave packet is given by the relation $\Delta v \, \Delta t \geq 1$ and Shannon taught us that the amount of information is proportional this product.